\documentclass[aps,prd,12pt,onecolumn,nofootinbib,superscriptaddress,longbibliography]{revtex4-2}
\usepackage{amsmath,amssymb,bm,booktabs,graphicx,hyperref,microtype,mathrsfs}
\hypersetup{hidelinks,pdftitle={Primordial Graviton Entanglement in Quasicrystalline Inflation},pdfauthor={Jiro Soda}}
\allowdisplaybreaks
\newcommand{\mpl}{M_{\rm Pl}}
\newcommand{\mpv}{M_{\rm PV}}
\newcommand{\dd}{\mathrm d}
\newcommand{\bk}{\bm k}
\newcommand{\bp}{\bm p}
\newcommand{\bq}{\bm q}
\newcommand{\bG}{\bm G}
\newcommand{\bn}{\bm n}
\newcommand{\cA}{\mathcal A}
\newcommand{\cB}{\mathcal B}
\newcommand{\cP}{\mathcal P}
\newcommand{\cV}{\mathcal V}
\newcommand{\cS}{\mathcal S}
\newcommand{\intk}{\int\!\frac{\dd^3 k}{(2\pi)^3}}
\newcommand{\intp}{\int\!\frac{\dd^3 p}{(2\pi)^3}}

\newcommand{\ord}{\mathcal O}
\newcommand{\tr}{\operatorname{Tr}}

\makeatletter
\AtBeginDocument{\immediate\write\@auxout{\string\citation{apsControl}}}
\makeatother
\begin{document}
\begin{flushright}
\hspace{7cm}{ KOBE-COSMO-26-13}
\end{flushright}

\title{Primordial Graviton Entanglement in Quasicrystalline Inflation}
\author{Jiro Soda}
\affiliation{Department of Physics, Kobe University, Kobe 657-8501, Japan}

\begin{abstract}
We study the entanglement structure of gravitons in quasicrystalline inflation. Both parity-even and parity-odd interactions generate graviton pairs with nonzero total momentum. For opposite-helicity pairs with identical momenta, the parity-even contribution can be nonzero, whereas the parity-odd contribution vanishes. The transition from inflation to radiation domination produces additional conventional two-mode squeezing. We derive the resulting normal and anomalous covariances.
The initially nonzero reciprocal blocks of the full field--momentum
covariance remain nonzero under homogeneous evolution, although
individual components can vanish. We show that this homogeneous, helicity-preserving evolution acts unitarily within each helicity sector and therefore preserves the entanglement between the complete right- and left-helicity sectors. Nevertheless, it can eliminate entanglement in the reduced state of a co-propagating pair. We illustrate this distinction using a solvable four-mode model.
\end{abstract}
\maketitle

\section{Introduction}\label{sec:intro}
In a spatially homogeneous universe, a time-dependent background can
supply energy but cannot transfer net comoving momentum. Pair-creation
terms in a quadratic field Hamiltonian therefore couple modes with
opposite comoving momenta, $\bk$ and $-\bk$. This is the usual momentum
structure of inflationary graviton squeezing
\cite{Grishchuk1989,Grishchuk1990,Albrecht1994}.
The notions of particles and squeezing also depend on the mode basis
used to define annihilation operators
\cite{AgulloBongaRibes2022,KannoSoda2022}.
We specify our choice of mode basis explicitly.

In the quasicrystalline inflation model proposed in
Ref.~\cite{SodaQC2026}, the quasiperiodic medium can supply nonzero
comoving momentum through its reciprocal vectors $\bG_{\bn}$.
These vectors are constructed from six unit vectors $\bq_A$ directed
along the axes joining opposite vertices of an icosahedron.
The medium is described by six phase fields, and its application to
scalar perturbations was developed in the same work.
This quasiperiodic structure should be distinguished from discrete
point-group symmetry alone. The solid-inflation model of
Ref.~\cite{KangNicolis2016} uses icosahedral symmetry but does not contain
the six phase fields or the three additional phason directions of the
quasicrystalline medium.

A Fourier component of the coupling with wavevector $\bG_{\bn}$ can
create a pair with momenta $\bk$ and $\bG_{\bn}-\bk$.
The momentum pattern thus differs from that of conventional inflation:
the partner need not propagate in the opposite direction.
The question is therefore not only how many pairs are produced, but
also which combinations of momenta and helicities are allowed.
We address this question for two specified interactions.
The first is a parity-even scalar modulation of the graviton kinetic
term. The second is a parity-odd tensor operator containing three
spatial derivatives and one Levi-Civita tensor.
Both involve the same set of reciprocal vectors, but their helicity
weights differ.

Parity-sensitive primordial tensor perturbations have been studied
extensively in Chern--Simons and higher-curvature models, in theories
with odd-order spatial derivatives, and in gauge-field models
\cite{Lue1999,AlexanderMartin2005,SatohKannoSoda2008,SatohSoda2008,
TakahashiSoda2009,MaleknejadReview,ObataSoda2016,AokiEFT2026}.
These models retain the zero-total-momentum pairing rule when the
background is spatially homogeneous.
Spatially localized parity-sensitive media offer a closer analogy.
For example, gravitational-wave scattering by axion domain walls has
been studied in Chern--Simons gravity and Nieh--Yan gravity
\cite{KannoDomain2023,SodaTakeuchi2026}.
Here, by contrast, the spatially varying medium affects pair creation
during inflation rather than only the propagation of an incident wave.

In this paper, we study the entanglement structure of gravitons in
quasicrystalline inflation. The entanglement of the full squeezed state
generated by the quasiperiodic medium must be distinguished from that
of its one-pair component.
A given mode $\bk$ can couple to several partners
$\bG_{\bn}-\bk$, leading to a multimode correlation structure.
The interactions can also entangle modes of opposite helicity,
subject to helicity-dependent selection rules.

In addition to the squeezing generated by the quasiperiodic medium,
we must include the conventional squeezing of modes $\bk$ and $-\bk$
produced during the transition from inflation to radiation domination.
A nonzero correlation, however, does not by itself imply entanglement.
Whether a selected pair is entangled must be determined from its
reduced state.
We show that the subsequent homogeneous evolution factorizes into
unitary transformations acting separately within the right- and
left-helicity sectors.
It therefore preserves entanglement between the complete helicity
sectors, but can reduce or eliminate entanglement in the reduced state
of a selected pair.
We illustrate this distinction using a solvable four-mode model.

The paper is organized as follows.
Section~II introduces the quasicrystalline medium and its parity-even
and parity-odd profiles.
Section~III defines the two interactions between gravitons and the
quasicrystalline medium.
Section~IV derives the Bogoliubov transformations for finite-duration
interaction pulses.
Section~V discusses the directions of the created pairs and squeezing
between opposite-helicity modes with identical momenta.
Section~VI presents our results on the evolution of the
full multimode covariance from inflation to radiation domination.
Section~VII examines the implications for entanglement and uses
a solvable four-mode model to distinguish entanglement between
the complete helicity sectors from that of a reduced pair.
The final section presents our conclusions.
Technical details are given in the appendices.

\section{Quasicrystalline medium}\label{sec:medium}

In this section, we introduce the icosahedral geometry, the six phase fields, the parity-even and parity-odd profiles, and the background spacetime.

\subsection{Icosahedral geometry }\label{app:ico}
 Icosahedral geometry is characterized by six unit vectors
\begin{align}
 \bq_1&=\mathcal N^{-1}(0,1,\tau)\ ,\quad
 \bq_2=\mathcal N^{-1}(0,1,-\tau)\ ,\quad
 \bq_3=\mathcal N^{-1}(1,\tau,0)\ ,\nonumber\\
 \bq_4 &=\mathcal N^{-1}(1,-\tau,0)\ ,\quad
 \bq_5=\mathcal N^{-1}(\tau,0,1)\ ,\quad
 \bq_6=\mathcal N^{-1}(\tau,0,-1)\ ,
 \label{eq:q56}
\end{align}
where we used $\tau=(1+\sqrt5)/2$ and $\mathcal N=\sqrt{1+\tau^2}$.
The dyadic matrices satisfy \(\sum_A q_A^i q_A^j=2\delta^{ij}\). If $\sum_An_A\bq_A=0$, its three components imply
 all six integers vanish. This proves uniqueness of the integer label, not real-linear independence of six vectors in three dimensions.

For an icosahedral quasicrystal, reciprocal vectors can be written using $\bq_A$,
\begin{equation}
 \bG_{\bn}=K \sum_{A=1}^{6} n_A {\bf q}_A,\qquad {\bm n} =\{n_1, \cdots, n_6\}\in{\mathbb Z}^6
 ,
 \label{eq:crystal_module}
\end{equation}
where $K$ is a comoving inverse length.
This set is often called a rank-six reciprocal module: six independent integer labels are needed, although the momenta are three-dimensional. We usually call it the reciprocal set. Its twelve vectors $\pm K\bq_A$ form the first star. A star is an orbit under the point-group rotations. Higher integer order does not necessarily mean larger physical $|\bG_{\bn}|$.

In order to discuss the parity transformation, we need to construct the reflection normals from the twelve unit vertices $\mathcal V=\{\pm\bq_A\}$. Two vertices $\bm v,\bm w$ are adjacent when $\bm v\cdot\bm w=1/\sqrt5$. Normalize their sums:
\begin{equation}
 \bm\beta(\bm v,\bm w)=\frac{\bm v+\bm w}{|\bm v+\bm w|}.
 \label{eq:roots}
\end{equation}
The thirty edge-midpoint directions give fifteen opposite pairs. Choosing one member from each pair gives the normals. Changing an individual choice changes only the overall normalization sign. The reflection $I-2\bm\beta\bm\beta^T$ maps the vertex set to itself. Equivalently, the thirty roots are the coordinate axes with both signs and the cyclic permutations of $(\pm\tau^{-1},\pm\tau,\pm1)/2$. This makes the degree-fifteen construction reproducible without listing a fitted set of constants.

\subsection{Six phases}

In quasicrystalline inflation, we borrow the symmetry structure of quasicrystal and utilize an effective field theory approach.
There, the following six phases play a central role:
\begin{equation}
 \Theta_A(\bm x)=K\bq_A\cdot\bm x+\phi_A,
 \qquad \Theta_A\sim\Theta_A+2\pi.
 \label{eq:phases}
\end{equation}
Here, $\phi_A$ are constant phase offsets.   
The phase description follows the kinematics of quasicrystals \cite{Levine1984,Shechtman1984,Baggioli2020}. Three changes of phase are physical translations,
\begin{equation}
 \delta\phi_A=K\bq_A\cdot\bm u.
 \label{eq:phonon}
\end{equation}
Three independent changes obey $\sum_A\bq_A\delta\phi_A=0$ and are phason directions. The calculation below uses a fixed classical phase profile. It neither quantizes phason fluctuations nor assumes the finite-temperature diffusive dynamics of laboratory phasons. The microscopic formation and lifetime of the profile are separate inputs.

We do not average its orientation or offsets. The response functions have a specified transformation under the icosahedral group. A selected set of offsets need not preserve that entire group. The distinction is the same as the distinction between a symmetric action and a less symmetric state.

\subsection{The parity-even profile}
Write
\begin{equation}
 F_E(\bm x)=\sum_{\bn\ne0}h_{\bn}e^{i\bG_{\bn}\cdot\bm x},
 \qquad h_{-\bn}=h_{\bn}^*.
 \label{eq:Fe}
\end{equation}
A useful explicit response is
\begin{align}
 F_E(\Theta)=\frac{\exp[b\sum_A\cos\Theta_A]}{I_0(b)^6}-1 \ , \quad 
 h_{\bn}=c_{\bn}e^{i\bn\cdot\bm\phi}\ ,\quad
 c_{\bn}=\frac{\prod_A I_{n_A}(b)}{I_0(b)^6}\quad(\bn\ne0)\ .
 \label{eq:evenfourier}
\end{align}
The real parameter $b>0$ controls the harmonic weights.  The modified Bessel functions of the first kind enter through
\begin{equation}
 e^{b\cos\Theta}=\sum_{m\in\mathbb Z}I_m(b)e^{im\Theta}.
 \label{eq:besselidentity}
\end{equation}
Multiplying six such series proves Eq.~\eqref{eq:evenfourier}. It does not force the average over each finite patch to vanish. When $b\ll1$, $I_1/I_0\simeq b/2$, so low integer orders dominate. For $b>0$, every coefficient $c_{\bn}$ in Eq.~\eqref{eq:evenfourier} is positive.

\subsection{The parity-odd profile}
The second source profile is built from
\begin{equation}
 V_i(\Theta)=\sum_Aq_{Ai}\sin\Theta_A.
 \label{eq:phasevector}
\end{equation}
The full icosahedral reflection group has fifteen reflection planes. Select one unit normal $\bm\beta_r$ from each opposite pair and set
\begin{equation}
 \Delta_{15}(\bm V)=\prod_{r=1}^{15}(\bm\beta_r\cdot\bm V).
 \label{eq:discriminant}
\end{equation}
The normals can be constructed directly from the vertices. A reflection-odd polynomial must vanish on each reflection plane, so it contains all fifteen linear factors. This explains the degree fifteen. The product is invariant under proper icosahedral rotations and changes sign under improper ones \cite{Humphreys1990}.

Use the real periodic response
\begin{equation}
 \chi(\Theta)=\frac{\Delta_{15}[\bm V(\Theta)]}{\chi_{\rm norm}}
 \exp\!\left[b\sum_A\cos\Theta_A\right],
 \label{eq:oddprofile}
\end{equation}
where the real nonzero constant $\chi_{\rm norm}$ sets its normalization. Since $\bm V(-\Theta)=-\bm V(\Theta)$,
\begin{equation}
 \chi(-\Theta)=-\chi(\Theta),\qquad
 \int\frac{\dd^6\Theta}{(2\pi)^6}\chi(\Theta)=0.
 \label{eq:oddness}
\end{equation}
With one Fourier convention for both profiles,
\begin{align}
 \chi(\bm x)=\sum_{\bn\ne0}j_{\bn}e^{i\bG_{\bn}\cdot\bm x}\ ,\quad
 j_{-\bn}=j_{\bn}^* \ ,\quad
 j_{\bn}=d_{\bn}e^{i\bn\cdot\bm\phi}\ ,\quad
 d_{-\bn}=d_{\bn}^*=-d_{\bn}\ .
 \label{eq:oddcoeff}
\end{align}
Thus, the coefficients $d_{\bn}$ are purely imaginary and $j_{\bm0}=0$. The zero coefficient is independent of the offsets. It equals the infinite-volume spatial mean of the ideal irrational profile, but not necessarily the mean in a finite coherent region.

The odd coefficients are not the Bessel products in Eq.~\eqref{eq:evenfourier}. The discriminant supplies additional coefficients and can make whole orbits vanish. In particular, the first-star coefficients of the explicit response \eqref{eq:oddprofile} vanish. Let \(\mathbf e_A\) denote the standard basis vector in \(\mathbb Z^6\), with \((\mathbf e_A)_B=\delta_{AB}\). A proper icosahedral rotation maps \(\mathbf e_A\) to \(-\mathbf e_A\) in phase-label space, so invariance gives \(d_{\mathbf e_A}=d_{-\mathbf e_A}\).  Oddness gives the opposite equality. Hence
\begin{equation}
 d_{\pm\bm e_A}=0.
 \label{eq:extinction}
\end{equation}
Coefficients at higher integer orders need not vanish.

\subsection{Parity symmetry and the background stress}
The words ``even'' and ``odd'' label the tensor operators used below. If $\chi$ transforms as a pseudoscalar, its product with a parity-odd tensor operator is a scalar in the full action. A chosen background may then break parity, but it need not do so for every phase choice. For example, $\chi(-\bm x)=-\chi(\bm x)$ is compatible with parity invariance of a pseudoscalar profile about that origin. A helicity sign change at fixed momenta is therefore not, by itself, a test of the parity of the whole background. Parity also reverses the momenta and transforms the coefficient. We keep these statements separate.

Likewise, fixed reciprocal directions already carry directional information. Statistical anisotropy is a property of the full covariance, including peak locations as well as their weights. Translation breaking gives the nonzero momentum-transfer rule; it is not identical to rotation breaking. No orientation average is used to separate them.

A simple background-compatible spectator is
\begin{equation}
 S_{\rm sp}=-\frac12\int\dd^4x\sqrt{-g}\,f^2(\varphi)
 \sum_Ag^{\mu\nu}\partial_\mu\Theta_A\partial_\nu\Theta_A,
 \label{eq:spectator}
\end{equation}
where $\varphi$ is the inflaton. The spectator equation is $\nabla_\mu(f^2\partial^\mu\Theta_A)=0$. On Eq.~\eqref{eq:phases}, the temporal derivative vanishes and the spatial gradients are constant, so the equation is satisfied. Thus, the linear profile solves its phase equations. 
Its stress is
\begin{equation}
 T_{\mu\nu}=f^2\sum_A\partial_\mu\Theta_A\partial_\nu\Theta_A
 -\tfrac12g_{\mu\nu}f^2\sum_A(\partial\Theta_A)^2.
 \label{eq:stressfull}
\end{equation}
The identity $\sum_Aq_A^iq_A^j=2\delta^{ij}$ gives
 $\sum_A(\partial\Theta_A)^2=6K^2/a^2$.
 Hence, Eq.~\eqref{eq:stressfull} reads
\begin{equation}
 \rho_{\rm sp}=3f^2K^2/a^2,\qquad p_{\rm sp}=-f^2K^2/a^2.
 \label{eq:stress}
\end{equation}
This stress is isotropic and independent of the offsets. We assume $f^2>0$ and $\Omega_{\rm sp}\ll\epsilon_H$, where $\Omega_{\rm sp}=\rho_{\rm sp}/(3\mpl^2H^2)$ and $\epsilon_H=-\dot H/H^2$. 
Here, $\rho_{\rm sp}, \mpl$, $H$ are the energy density of the spectator field, the Planck mass and the Hubble parameter.  This establishes compatibility with an almost de Sitter background, not the formation of a relativistic quasicrystal. We work to leading order in the spectator energy fraction and neglect its homogeneous corrections to tensor propagation. The tensor interactions below start at quadratic order in perturbations and give no background tensor tadpole.

\section{Gravitons in quasicrystalline inflation}\label{sec:action}

In this section, we consider gravitons in quasicrystalline inflation. There are parity even and odd interactions.

The gravitons are described by the metric
\begin{eqnarray}
  ds^2=a^2(\eta)[-d\eta^2+(e^\gamma)_{ij}dx^idx^j] 
  \ ,  
\end{eqnarray}
where $a$ is the scale factor. 
Using this definition, we can write down the quadratic action for the gravitons
\begin{align}
 S_T&=S_0+S_E+S_O,\label{eq:Stotal}\\
 S_0&=\frac{\mpl^2}{8}\int\dd\eta\,\dd^3x\,a^2
 [\gamma_{ij}'\gamma_{ij}'-\partial_m\gamma_{ij}\partial_m\gamma_{ij}],\label{eq:S0}\\
 S_E&=\frac{\mpl^2}{8}\int\dd\eta\,\dd^3x\,a^2
 \varepsilon_E s_E(\eta)F_E(\bm x)\gamma_{ij}'\gamma_{ij}',\label{eq:SE}\\
 S_O&=\frac{\mpl^2}{8\mpv}\int\dd\eta\,\dd^3x\,a c_O(\eta)\chi(\bm x)
 \epsilon^{ijk}(\partial^2\gamma_{i\ell})\partial_j\gamma_{k\ell}.
 \label{eq:SO}
\end{align}
Here $\partial_i\gamma_{ij}=0$, $\gamma_{ii}=0$, and $\epsilon^{123}=+1$. All spatial derivatives are comoving and primes denote conformal-time derivatives. 
The parity-even coupling is controlled by a constant $\varepsilon_E$. The dimensionless function $s_E(\eta)$ describes a pulse interaction. The parameter \(M_{\rm PV}\) is the mass scale of the parity-odd term. The function \(c_O(\eta)\) is real and dimensionless. The free and kinetic terms are motivated by the inflationary effective theory; parity-odd spatial operators occur in spatially covariant gravity \cite{Cheung2008,DeFeliceTsujikawa2015,Nishizawa2018,GaoHong2020,SongGao2026}. We take Eqs.~\eqref{eq:Stotal}--\eqref{eq:SO} as the specified tensor model. We do not claim a unique nonlinear completion or absence of scalar--tensor mixing in every completion.

The kinetic coefficient must satisfy
\begin{equation}
 Q_T=1+\varepsilon_Es_E F_E>0,
 \qquad |\varepsilon_Es_EF_E|\ll1.
 \label{eq:kineticcontrol}
\end{equation}
When the coefficients vary little over a wavelength, the local tensor frequency is
\begin{equation}
\begin{aligned}
 \omega_\lambda^2 \simeq
 \frac{k^2+\sigma_\lambda c_O\chi\,k^3/(a\mpv)}{Q_T}\ ,  \quad
 \sigma_R =+1,\qquad\sigma_L=-1.
\end{aligned}
 \label{eq:dispersion}
\end{equation}
The sign follows our plus sign for $S_O$ and the polarization convention below. Changing that action sign reverses the odd amplitudes. A local sufficient smallness condition is $|c_O\chi|k/(a\mpv)\ll1$. The odd term adds no time derivative, so it does not create a new tensor kinetic degree of freedom.

A spatially varying coefficient need not commute with the transverse-traceless projection. It is therefore safer to perform the Legendre transform in a perturbative manner. It gives $H_{E,I}=-L_{E,I}$ at first order. Since $S_O$ has no velocity dependence, we also get $H_{O,I}=-L_{O,I}$. Thus
\begin{align}
 H_I={}&-\frac{\mpl^2}{8}\int\dd^3x\,a^2\varepsilon_Es_EF_E\gamma_{ij,I}'\gamma_{ij,I}'
 \nonumber\\[-1mm]
 &-\frac{\mpl^2}{8\mpv}\int\dd^3x\,a c_O\chi\,
 \epsilon^{ijk}(\partial^2\gamma_{i\ell,I})\partial_j\gamma_{k\ell,I}
 +\ord(\varepsilon_E^2).
 \label{eq:HI}
\end{align}
 Products of the two couplings are second order in amplitudes, but interference between their first-order pair amplitudes contributes to the leading pair probability.

Normalize the oscillators and tensors by
\begin{equation}
\begin{aligned}
 \relax [a_{\bk\lambda},a_{\bp\lambda'}^\dagger]=(2\pi)^3\delta_{\lambda\lambda'}\delta^3(\bk-\bp)\ ,\qquad
 e_{ij}^\lambda(\hat\bk)e_{ij}^{\lambda'*}(\hat\bk)=\delta_{\lambda\lambda'}\ .
\end{aligned}
 \label{eq:norm}
\end{equation}
The interaction-picture field is
\begin{equation}
\begin{aligned}
 \gamma_{ij,I}(\eta,\bm x)=\sum_\lambda\intk
 \bigl[e_{ij}^\lambda(\hat\bk)u_k a_{\bk\lambda}e^{i\bk\cdot\bm x}
 +e_{ij}^{\lambda*}(\hat\bk)u_k^*a_{\bk\lambda}^\dagger e^{-i\bk\cdot\bm x}\bigr]\ .
\end{aligned}
 \label{eq:field}
\end{equation}
With $v_\lambda=\mpl a\gamma_\lambda/2$, the free action becomes canonical after one integration by parts. Its equation is
\begin{equation}
 v_k''+\left(k^2-\frac{a''}{a}\right)v_k=0.
 \label{eq:canonicalmode}
\end{equation}
For $a=-1/(H\eta)$, choose
\begin{align}
 u_k(\eta)=\frac{\sqrt2H}{\mpl k^{3/2}}(1+ik\eta)e^{-ik\eta}\ ,\qquad
 v_k(\eta)=-\frac{i}{\sqrt{2k}}\left(1-\frac{i}{k\eta}\right)e^{-ik\eta}.
 \label{eq:vmode}
\end{align}
This has $u_k(0)>0$ and $v_kv_k^{*\prime}-v_k'v_k^*=i$. The constant phase $-i$ in $v_k$ is kept in the radiation matching. The notation $u_k(0)$ denotes the formal de Sitter late-time normalization; our interaction ends at $\eta_s<0$.

In a transverse orthonormal frame $(\bm t,\bm y,\hat\bk)$ with $\bm y=\hat\bk\times\bm t$, define
\begin{equation}
 \bm m_\sigma=\frac{\bm t+i\sigma\bm y}{\sqrt2},\qquad
 e_{ij}^{\sigma}=m_{\sigma i}m_{\sigma j}.
 \label{eq:circular}
\end{equation}
Then
\begin{equation}
 i\epsilon^{ijk}\hat k_j e_{k\ell}^{\sigma}=\sigma e_{i\ell}^{\sigma}.
 \label{eq:helicityidentity}
\end{equation}
The helicity of the graviton is $2\sigma$; $\sigma=\pm1$ is only its sign. Complex-conjugated polarization tensors have the opposite eigenvalue in Eq.~\eqref{eq:helicityidentity}. This fact fixes the creation-vertex signs.

\section{Bogoliubov maps }\label{sec:bogoliubov}

Let $U_I$ be the time-evolution operator generated by Eq.~\eqref{eq:HI}. Define
\begin{equation}
 d_a=U_I^\dagger a_aU_I
 =a_a+i\int\dd\eta\,[H_I(\eta),a_a]+\ord(\varepsilon^2),
 \label{eq:operator}
\end{equation}
where $a$ now labels a momentum and helicity. 
Repeated mode labels imply a helicity sum and a momentum integral with measure \(d^3k/(2\pi)^3\).
The state is $U_I|0\rangle$. Taking moments of the $d_a$ in $|0\rangle$ gives the same result as taking moments of the original oscillators in that state. The $d_a$ use the same free de Sitter mode basis after the pulse; they are not yet radiation-era particle operators.

A general quadratic Hamiltonian has the form
\begin{equation}
\begin{aligned}
 H_I=\Omega_{ab}a_a^\dagger a_b+\tfrac12
 (\Xi_{ab}a_a^\dagger a_b^\dagger+\Xi_{ab}^*a_ba_a)\ ,\qquad
 \Omega^\dagger=\Omega,\qquad\Xi^T=\Xi.
\end{aligned}
 \label{eq:generalH}
\end{equation}
The commutators needed in Eq.~\eqref{eq:operator} are
\begin{align}
 [a_b^\dagger a_c,a_a]=-\delta_{ab}a_c \ ,\quad
 [a_b^\dagger a_c^\dagger,a_a]=-\delta_{ab}a_c^\dagger-\delta_{ac}a_b^\dagger.
 \label{eq:comm2}
\end{align}
The two terms in the second commutator
 are equal after contraction with the symmetric $\Xi_{bc}$. Hence
\begin{equation}
\begin{aligned}
 d_a =(\delta_{ab}+\cA_{ab})a_b+\cB_{ab}a_b^\dagger+\ord(\varepsilon^2)\ ,\qquad
 \cA =-i\int\Omega\,\dd\eta,\qquad\cB=-i\int\Xi\,\dd\eta\ .
\end{aligned}
 \label{eq:bog}
\end{equation}
The block $\cA$ mixes annihilation operators and describes passive, number-conserving scattering. The block $\cB$ mixes annihilation and creation operators and describes active pair creation. A time-dependent quadratic action produces a Gaussian state, but the exact finite-strength map generally requires time ordering.

For an exact map $d=\alpha a+\beta a^\dagger$, preserving both canonical commutators gives
\begin{equation}
 \alpha\alpha^\dagger-\beta\beta^\dagger=I,
 \qquad \alpha\beta^T-\beta\alpha^T=0.
 \label{eq:canonicalconditions}
\end{equation}
Putting $\alpha=I+\cA+\ord(\varepsilon^2)$ and $\beta=\cB+\ord(\varepsilon^2)$ yields
\begin{equation}
 \cA+\cA^\dagger=0,\qquad\cB^T=\cB.
 \label{eq:firstcanonical}
\end{equation}
These are useful checks on the kernels derived next. 

\subsection{Momentum supplied by the spatial profiles}
Both profiles use $e^{+i\bG_{\bn}\cdot\bm x}$. Two creation waves in Eq.~\eqref{eq:field} contribute $e^{-i(\bk+\bp)\cdot\bm x}$. A creation and an annihilation wave contribute $e^{-i(\bk-\bp)\cdot\bm x}$. Thus
\begin{align}
 \int\dd^3x\,e^{i(\bG_{\bn}-\bk-\bp)\cdot\bm x}
 &=(2\pi)^3\delta^3(\bk+\bp-\bG_{\bn}),\label{eq:pairdelta}\\
 \int\dd^3x\,e^{i(\bG_{\bn}-\bk+\bp)\cdot\bm x}
 &=(2\pi)^3\delta^3(\bk-\bp-\bG_{\bn}).\label{eq:mixdelta}
\end{align}
It follows that the total first-order blocks have support
\begin{align}
 \cB_{\lambda\lambda'}(\bk,\bp)&=(2\pi)^3\sum_{\bn\ne0}
 \delta^3(\bk+\bp-\bG_{\bn})\cB_{\bn}^{\lambda\lambda'}(\bk,\bp),\label{eq:Bsupport}\\
 \cA_{\lambda\lambda'}(\bk,\bp)&=(2\pi)^3\sum_{\bn\ne0}
 \delta^3(\bk-\bp-\bG_{\bn})\cA_{\bn}^{\lambda\lambda'}(\bk,\bp).
 \label{eq:Asupport}
\end{align}
The coefficients of the two interactions add at this order. 

\subsection{The parity-even Bogoliubov coefficients}
Let us define the polarization overlap
\begin{equation}
 \cP^+_{\lambda\lambda'}=e_{ij}^{\lambda*}(\hat\bk)e_{ij}^{\lambda'*}(\hat\bp),
 \qquad
 \cP^-_{\lambda\lambda'}=e_{ij}^{\lambda*}(\hat\bk)e_{ij}^{\lambda'}(\hat\bp).
 \label{eq:overlaps}
\end{equation}
The superscripts indicate the active and passive contraction, not the linear $+$ polarization. Substitution of the first term in Eq.~\eqref{eq:HI} gives
\begin{align}
 \cB_{E,\bn}^{\lambda\lambda'}&=
 \frac{i\varepsilon_E\mpl^2}{4}h_{\bn}\cP^+_{\lambda\lambda'}
 \int\dd\eta\,a^2s_Eu_k^{\prime*}u_p^{\prime*},\label{eq:BEint}\\
 \cA_{E,\bn}^{\lambda\lambda'}&=
 \frac{i\varepsilon_E\mpl^2}{4}h_{\bn}\cP^-_{\lambda\lambda'}
 \int\dd\eta\,a^2s_Eu_k^{\prime*}u_p'.\label{eq:AEint}
\end{align}
The $1/4$ follows from the $1/8$ in the action and the two oscillator contractions in Eq.~\eqref{eq:comm2}. The same factor occurs in the passive term because the two placements of its creation operator are equal.

\subsection{The parity odd Bogoliubov coefficients}
For the two-creation term in $S_O$, the spatial derivatives give $(-k^2)(-ip_j)$, not $(-k^2)(+ip_j)$. The symmetrized creation vertex is therefore
\begin{align}
 \cV^+_{\lambda\lambda'}(\bk,\bp)=\frac{i}{2}\epsilon^{ijk}
 \bigl[k^2p_j e_{i\ell}^{\lambda*}(\hat\bk)e_{k\ell}^{\lambda'*}(\hat\bp)
 +p^2k_j e_{i\ell}^{\lambda'*}(\hat\bp)e_{k\ell}^{\lambda*}(\hat\bk)\bigr].
 \label{eq:Vplus}
\end{align}
For one creation leg at $\bk$ and one annihilation leg at $\bp$, the corresponding vertex is
\begin{align}
 \cV^-_{\lambda\lambda'}(\bk,\bp)=\frac{i}{2}\epsilon^{ijk}
 \bigl[-k^2p_j e_{i\ell}^{\lambda*}(\hat\bk)e_{k\ell}^{\lambda'}(\hat\bp)
 +p^2k_j e_{i\ell}^{\lambda'}(\hat\bp)e_{k\ell}^{\lambda*}(\hat\bk)\bigr].
 \label{eq:Vminus}
\end{align}
These expressions explicitly fix the complex conjugations. Using Eq.~\eqref{eq:helicityidentity} separately on each differentiated tensor gives the useful exact formulas
\begin{equation}
 \displaystyle
 \cV^\pm_{\lambda\lambda'}(\bk,\bp)=
 -\frac{kp}{2}(k\sigma_{\lambda'}+p\sigma_\lambda)\cP^\pm_{\lambda\lambda'}(\bk,\bp).
 \label{eq:Vsimple}
\end{equation}
For example, $i\epsilon^{ijk}p_je_{k\ell}^{\lambda'*}=-\sigma_{\lambda'}p\,e_{i\ell}^{\lambda'*}$. The first term of Eq.~\eqref{eq:Vplus} is then $-k^2p\sigma_{\lambda'}\cP^+/2$. The other term gives $-p^2k\sigma_\lambda\cP^+/2$. Appendix~\ref{app:polarization} provides a direct contraction check for general directions.

The odd blocks are
\begin{align}
 \cB_{O,\bn}^{\lambda\lambda'}&=
 \frac{i\mpl^2}{4\mpv}j_{\bn}\cV^+_{\lambda\lambda'}
 \int\dd\eta\,a c_Ou_k^*u_p^*,\label{eq:BOint}\\
 \cA_{O,\bn}^{\lambda\lambda'}&=
 \frac{i\mpl^2}{4\mpv}j_{\bn}\cV^-_{\lambda\lambda'}
 \int\dd\eta\,a c_Ou_k^*u_p.\label{eq:AOint}
\end{align}
The active vertex is symmetric under exchange of the complete mode labels. The passive vertex satisfies $\cV^-_{\lambda\lambda'}(\bk,\bp)=\cV_{\lambda'\lambda}^{-*}(\bp,\bk)$. Together with real time profiles and $j_{-\bn}=j_{\bn}^*$, these properties imply Eq.~\eqref{eq:firstcanonical}. The total blocks are $\cB=\cB_E+\cB_O$ and $\cA=\cA_E+\cA_O$.

\subsection{Finite pulses and time integrals}\label{subsec:pulses}
To define the state after the interaction, we use pulses that decay exponentially as \(\eta\to-\infty\) and vanish for \(\eta\ge\eta_s\). A particle-state calculation needs a finite interaction period. We let both interactions end at $\eta_s<\eta_1<0$, before the radiation transition. Define a family of real pulses
\begin{equation}
 s_m(\eta)=
 \begin{cases}
 \displaystyle\frac1{m!}\left(\frac{\eta_s-\eta}{\Delta}\right)^m
 e^{(\eta-\eta_s)/\Delta},&\eta<\eta_s,\\[1mm]
 0,&\eta\ge\eta_s,
 \end{cases}
\quad
 \label{eq:pulse}
\end{equation}
Here $\Delta>0$ and $\eta_s<\eta_1<0$: the pulse ends before the background transition. The pulse and its first $m-1$ derivatives vanish at its endpoint. We choose
\begin{equation}
 s_E=s_3,\qquad c_O(\eta)=\varepsilon_O\frac{a_s}{a(\eta)}s_4(\eta),
 \qquad a_s=a(\eta_s).
 \label{eq:pulsechoices}
\end{equation}
The even pulse is the cubic pulse used in the kinetic model. The odd profile is a new finite-time benchmark for the same three-derivative operator. The factor $a_s/a$ makes $a c_O=\varepsilon_Oa_s s_4$, so its time integrals are elementary. The extra power in $s_4$ gives a sufficiently soft ultraviolet tail. No power-law coupling is extended to $\eta=0$ in the produced-state calculation.

Put $y=(\eta_s-\eta)/\Delta$. The Fourier transform is
\begin{equation}
\begin{aligned}
 \cS_m(\omega)=\int\dd\eta\,s_m(\eta)e^{i\omega\eta}
 =\frac{\Delta e^{i\omega\eta_s}}{m!}
 \int_0^\infty\dd y\,y^m e^{-(1+i\omega\Delta)y}
 =\frac{\Delta e^{i\omega\eta_s}}{(1+i\omega\Delta)^{m+1}}.
\end{aligned}
 \label{eq:Sm}
\end{equation}
This uses $\int_0^\infty y^m e^{-zy}\dd y=m!/z^{m+1}$ for $\operatorname{Re}z>0$.

For the even interaction,
\begin{equation}
 a^2u_k^{\prime*}u_p^{\prime*}=\frac{2\sqrt{kp}}{\mpl^2}e^{i(k+p)\eta},
 \qquad
 a^2u_k^{\prime*}u_p'=\frac{2\sqrt{kp}}{\mpl^2}e^{i(k-p)\eta}.
 \label{eq:derivproducts}
\end{equation}
The result is
\begin{align}
 \cB_{E,\bn}^{\lambda\lambda'}&=
 \frac{i\varepsilon_Eh_{\bn}}2\sqrt{kp}\,\cS_3(k+p)\cP^+_{\lambda\lambda'},\label{eq:BEclosed}\\
 \cA_{E,\bn}^{\lambda\lambda'}&=
 \frac{i\varepsilon_Eh_{\bn}}2\sqrt{kp}\,\cS_3(k-p)\cP^-_{\lambda\lambda'}.
 \label{eq:AEclosed}
\end{align}
For the odd interaction, let $\omega=k+p$, $\delta=k-p$, and $q=kp$. Differentiation of a Fourier transform gives $\int\eta s_me^{i\omega\eta}\dd\eta=-i\cS_m'(\omega)$ and $\int\eta^2s_me^{i\omega\eta}\dd\eta=-\cS_m''(\omega)$. Since
\begin{align}
 (1-ik\eta)(1-ip\eta)&=1-i\omega\eta-q\eta^2,\label{eq:polyplus}\\
 (1-ik\eta)(1+ip\eta)&=1-i\delta\eta+q\eta^2,
 \label{eq:polyminus}
\end{align}
define
\begin{align}
 \mathcal F_+(k,p)&=\cS_4(\omega)-\omega\cS_4'(\omega)+q\cS_4''(\omega),\label{eq:Fplus}\\
 \mathcal F_-(k,p)&=\cS_4(\delta)-\delta\cS_4'(\delta)-q\cS_4''(\delta).
 \label{eq:Fminus}
\end{align}
Primes on $\cS_m$ denote differentiation with respect to its frequency argument, not conformal time. The closed odd kernels are
\begin{align}
 \cB_{O,\bn}^{\lambda\lambda'}&=
 \frac{i\varepsilon_Oa_sH^2j_{\bn}}{2\mpv(kp)^{3/2}}
 \cV^+_{\lambda\lambda'}\mathcal F_+(k,p),\label{eq:BOclosed}\\
 \cA_{O,\bn}^{\lambda\lambda'}&=
 \frac{i\varepsilon_Oa_sH^2j_{\bn}}{2\mpv(kp)^{3/2}}
 \cV^-_{\lambda\lambda'}\mathcal F_-(k,p).
 \label{eq:AOclosed}
\end{align}
All four expressions are dimensionless pair or scattering amplitudes after the reciprocal delta function has been removed. Their momenta remain comoving; physical frequencies are $k/a$ and $p/a$.

\section{Pair directions and helicity selection}\label{sec:geometry}

For one reciprocal vector, define
\begin{equation}
 R_{\bG}(\bk)=\bG-\bk,\qquad R_{\bG}^2(\bk)=\bk.
 \label{eq:reflection}
\end{equation}
The midpoint is $(\bk+\bp)/2=\bG/2$. This is reflection through a point in momentum space, not reflection from a physical mirror. Two such maps give
\begin{equation}
 R_{\bG_r}[R_{\bG_s}(\bk)]=\bk+\bG_r-\bG_s.
 \label{eq:composition}
\end{equation}
Thus the possible links share one reciprocal structure. Equation (\ref{eq:composition}) describes connectivity in momentum space. Direct pair-creation links obey Eq. (38) and appear at first order. The perturbative order of correlations generated along a multistep path must be determined from the evolution map.

If $\vartheta$ is the opening angle, then
\begin{equation}
 G^2=k^2+p^2+2kp\cos\vartheta,
 \qquad
 \cos\vartheta=\frac{G^2-k^2-p^2}{2kp}.
 \label{eq:triangle}
\end{equation}
The allowed range is $|k-p|\le G\le k+p$. For equal magnitudes,
\begin{equation}
 G=2k\cos(\vartheta/2).
 \label{eq:equalangle}
\end{equation}
The homogeneous limit gives $\vartheta=\pi$. The fixed point $\bk=\bp=\bG/2$ gives $\vartheta=0$. The latter requires finite $G/k$ and cannot be obtained by expanding around a nearly opposite pair.

Choose transverse frames in the plane of $\bk$ and $\bp$. A direct vector contraction, shown in Appendix~\ref{app:polarization}, gives
\begin{equation}
 \cP^+_{\lambda\lambda'}(\vartheta)
 =\frac{(1-\sigma_\lambda\sigma_{\lambda'}\cos\vartheta)^2}{4}.
 \label{eq:angularP}
\end{equation}
Other frames attach helicity phases but leave its absolute value unchanged. At opposite directions only equal helicities have nonzero overlap. At the same direction only opposite helicities do:
\begin{equation}
 \cP^+_{RR}(0)=\cP^+_{LL}(0)=0,
 \qquad\cP^+_{RL}(0)=\cP^+_{LR}(0)=1.
 \label{eq:collinearP}
\end{equation}
Near co-propagation, the equal-helicity amplitude is $\vartheta^4/16+\ord(\vartheta^6)$ and the opposite-helicity amplitude is $1-\vartheta^2/2+\ord(\vartheta^4)$. These factors are amplitudes; probabilities contain their squares.

Combining Eqs.~\eqref{eq:Vsimple} and \eqref{eq:angularP} shows that the odd operator uses the same angular overlap, multiplied by a momentum-weighted helicity sign. For a nearly opposite pair with $k\simeq p$, it gives $\cV^+_{\lambda\lambda'}\simeq-\sigma_\lambda k^3\delta_{\lambda\lambda'}$. The leading $RR$ and $LL$ amplitudes therefore have opposite signs in our convention. Their separate probabilities are equal in the absence of interference with another interaction.

Let $\bk=k\hat{\bm n}$ and $\bp=p\hat{\bm n}$. The reciprocal condition is $\bG=(k+p)\hat{\bm n}$. Equation~\eqref{eq:Vsimple} gives
\begin{align}
 \cV^+_{RR}=\cV^+_{LL}=0\ , \quad
 \cV^+_{RL}=-\tfrac12kp(p-k)\ ,\quad
 \cV^+_{LR}=+\tfrac12kp(p-k)\ .\label{eq:collLR}
\end{align}
The even kernel remains nonzero at $k=p$, whereas the odd kernel vanishes there. The latter is a zero of the chosen operator, not a universal result for every parity-odd tensor coupling.

The cancellation also has a simple state interpretation. For unequal frequency modes, the odd one-pair term is proportional to
\begin{equation}
 a_{kL}^\dagger a_{pR}^\dagger-a_{kR}^\dagger a_{pL}^\dagger.
 \label{eq:oddpaircreation}
\end{equation}
As the two complete spatial and frequency profiles become identical, the two products commute and coincide. Their difference vanishes. The helicity part is antisymmetric, and its frequency-dependent coefficient changes sign under $k\leftrightarrow p$. The full bosonic amplitude is symmetric under exchange of the two complete labels.

\section{ Evolution and Survival of Graviton Correlations }
\label{sec:radiation}

We now address the central question of how the graviton correlations
generated by the quasicrystalline medium evolve after the interactions
end. We derive their late-time covariance and show that homogeneous
evolution cannot erase its initially nonzero reciprocal blocks.
This result concerns the full covariance; entanglement in a selected
subsystem requires a separate analysis, which we give in Sec.~VII.

All maps in this section use the total blocks $\cA=\cA_E+\cA_O$ and $\cB=\cB_E+\cB_O$. The propagation result does not depend on which interaction created the state.
After both pulses have vanished, the tensor equation again has spatially homogeneous coefficients. This does not set its initial correlation function to a homogeneous one. The medium can disappear while the initial data retain its spatial pattern. Closely related inhomogeneous primordial covariances and their later CMB transfer have been studied in Ref.~\cite{Kothari2016}.

The safest statement uses both the field and its canonical momentum. After switch-off, define $X_{\bk}=(\gamma_{\bk},\Pi_{\bk})^T$, where $\Pi_{\bk}=\mpl^2a^2\gamma'_{\bk}/4$ for a normalized helicity component. Homogeneous linear evolution has the form
\begin{equation}
 X_{\bk}(\eta)=T_k(\eta,\eta_s)X_{\bk}(\eta_s),\qquad \det T_k=1.
 \label{eq:transfer}
\end{equation}
We define \(\delta \mathscr C=\mathscr C-{\mathscr C}_{\mathrm{hom}}\), where \({\mathscr C}_{\mathrm{hom}}\) is the covariance in the same background with both quasicrystalline interactions switched off.
For the symmetrized covariance $\mathscr C_{ij}(\bk,\bp)=\langle\{X_{\bk,i},X_{\bp,j}\}\rangle/2$,
\begin{equation}
 \delta\mathscr C(\bk,\bp;\eta)
 =T_k\,\delta\mathscr C(\bk,\bp;\eta_s)\,T_p^T.
 \label{eq:fullcovtransfer}
\end{equation}
Both matrices are invertible. A nonzero full covariance block at $\bk+\bp=\bG_{\bn}$ cannot be mapped to the zero block by this evolution. Its field--field entry can vanish at a particular time, with information in the field--momentum or momentum--momentum entries. Thus it is the complete covariance support, not every individual measured amplitude, that is protected.

\subsection{De Sitter--radiation matching}
To display the particle-basis change, consider the continuous background
\begin{equation}
 a(\eta)=\begin{cases}
 -1/(H\eta),&\eta\le\eta_1<0,\\
 (\eta-2\eta_1)/(H\eta_1^2),&\eta\ge\eta_1.
 \end{cases}
 \label{eq:background}
\end{equation}
Both $a$ and $a'$ are continuous. The step in $a''/a$ has no delta function, so $v_k$ and $v_k'$ are continuous. In radiation domination $a''=0$ and the canonical modes are free oscillators. We choose
\begin{equation}
 w_k(\eta)=-\frac{i}{\sqrt{2k}}e^{-ik(\eta-2\eta_1)},\qquad
 v_k^{\rm I}=\alpha_k^0w_k+\beta_k^{0*}w_k^*.
 \label{eq:radmodes}
\end{equation}
The constant phase matches our convention in Eq.~\eqref{eq:vmode}. The radiation annihilation operators are
\begin{equation}
 b_{\bk\lambda}=\alpha_k^0 d_{\bk\lambda}+\beta_k^0d_{-\bk,\lambda}^\dagger.
 \label{eq:hommap}
\end{equation}
Solving the two matching equations, as shown in Appendix~\ref{app:match}, gives for $x=-k\eta_1>0$
\begin{equation}
\begin{aligned}
 \alpha_k^0=e^{2ix}\left(1+\frac{i}{x}-\frac1{2x^2}\right),\quad
 \beta_k^0=-\frac1{2x^2},\qquad |\alpha_k^0|^2-|\beta_k^0|^2=1.
\end{aligned}
 \label{eq:matchcoeff}
\end{equation}
The sign of $\beta_k^0$ depends on the stated mode phases. Its magnitude does not. The homogeneous occupation of the radiation mode for vacuum input is $|\beta_k^0|^2=1/(4x^4)$. It is large for modes outside the horizon at the transition and small for modes deep inside it. There is no further free canonical particle production during radiation domination itself. The coefficients encode the earlier expansion and its matching.

\subsection{Late-time covariance and momentum correlations}

We first combine the quasicrystal-induced transformation with the
subsequent homogeneous evolution. Substituting
Eq.~\eqref{eq:bog} into Eq.~\eqref{eq:hommap}, we obtain the
first-order corrections to the late-time Bogoliubov kernels:
\begin{align}
 \delta\alpha_{\rm late}(\bk,\bp)
 &=\alpha_k^0\cA(\bk,\bp)
   +\beta_k^0\cB^*(-\bk,\bp),
 \label{eq:totalalpha}\\
 \delta\beta_{\rm late}(\bk,\bp)
 &=\alpha_k^0\cB(\bk,\bp)
   +\beta_k^0\cA^*(-\bk,\bp).
 \label{eq:totalbeta}
\end{align}
Here $\cA$ and $\cB$ are the first-order mode-mixing
(passive) and pair-creation (active) kernels, respectively.
We suppress helicity indices throughout this subsection.
Products of kernels include the required helicity sums and
momentum integrals.

Writing
\[
 D_{\bk,\bp}=(2\pi)^3\delta^3(\bk-\bp),
\]
the zeroth-order kernels are
$\alpha_k^0D_{\bk,\bp}$ and $\beta_k^0D_{\bk,-\bp}$,
with helicity Kronecker deltas understood.
The two terms in Eq.~\eqref{eq:totalbeta} are supported at
$\bk+\bp=\bG_{\bn}$ and $\bk+\bp=-\bG_{\bn}$, respectively.
Because the reciprocal set contains both signs, the allowed
total pair momenta remain within the same set.
This support condition specifies where contributions can occur;
it does not require every allowed amplitude to be nonzero.

The pair-creation kernel $\beta_{\rm late}$ is not itself the
anomalous covariance. To see this, write the full transformation as
$b=\alpha_{\rm late}a+\beta_{\rm late}a^\dagger$, where $a$
annihilates the initial vacuum. The anomalous and normal
covariances are then
\begin{equation}
 \begin{aligned}
 M^b_{ij}
 &\equiv\langle b_i b_j\rangle
   =(\alpha_{\rm late}\beta_{\rm late}^{T})_{ij},\\
 N^b_{ij}
 &\equiv\langle b_i^\dagger b_j\rangle
   =\sum_l\beta_{{\rm late},il}^{*}
          \beta_{{\rm late},jl}.
 \end{aligned}
 \label{eq:NMlate}
\end{equation}
The indices include both momentum and helicity, and the sum over
$l$ includes momentum integration.
Both Eqs.~\eqref{eq:totalalpha} and \eqref{eq:totalbeta}
are therefore needed to calculate the late-time covariances.

In particular, a purely passive transformation leaves the initial
vacuum unchanged. Applying the same homogeneous evolution afterward
must therefore give the same state as in the absence of that passive
transformation. Its apparent first-order contributions cancel from
the covariances when both late-time kernels are included.
This cancellation also applies to the combined passive contributions
from the two interactions.

The covariance evolution can be written directly in terms of the
operators $d$ after the interaction pulses. Define
\[
 N^d_{\bk\bp}=\langle d_{\bk}^\dagger d_{\bp}\rangle,
 \qquad
 M^d_{\bk\bp}=\langle d_{\bk}d_{\bp}\rangle.
\]
Using Eq.~\eqref{eq:hommap}, we find
\begin{align}
 M^b_{\bk\bp}
 &=
 \alpha_k^0\alpha_p^0M^d_{\bk\bp}
 +\beta_k^0\beta_p^0M^{d*}_{-\bk,-\bp}
 +\alpha_k^0\beta_p^0
  [D_{\bk,-\bp}+N^d_{-\bp,\bk}]
 +\beta_k^0\alpha_p^0N^d_{-\bk,\bp},
 \label{eq:Mtransform}\\
 N^b_{\bk\bp}
 &=
 \alpha_k^{0*}\alpha_p^0N^d_{\bk\bp}
 +\beta_k^{0*}\beta_p^0
  [D_{\bk,\bp}+N^d_{-\bp,-\bk}]
 +\alpha_k^{0*}\beta_p^0M^{d*}_{\bk,-\bp}
 +\beta_k^{0*}\alpha_p^0M^d_{-\bk,\bp}.
 \label{eq:Ntransform}
\end{align}

For the two weak pulses acting on the initial vacuum,
\[
 M^d=\cB+O(\varepsilon^2),
 \qquad
 N^d=O(\varepsilon^2).
\]
Here $\varepsilon$ collectively denotes the small amplitudes of the
two pulses, and $O(\varepsilon^2)$ includes their mixed
second-order contributions.
The first relation also uses the first-order canonical condition
$\cB=\cB^T$.

In the absence of the modulation, the homogeneous covariances are
\[
 M^b_{0,\bk\bp}
 =\alpha_k^0\beta_p^0D_{\bk,-\bp},
 \qquad
 N^b_{0,\bk\bp}
 =\beta_k^{0*}\beta_p^0D_{\bk,\bp}.
\]
Subtracting these contributions gives
\begin{align}
 \delta M^b_{\bk\bp}
 &=
 \alpha_k^0\alpha_p^0\cB(\bk,\bp)
 +\beta_k^0\beta_p^0\cB^*(-\bk,-\bp)
 +O(\varepsilon^2),
 \label{eq:deltaM}\\
 \delta N^b_{\bk\bp}
 &=
 \alpha_k^{0*}\beta_p^0\cB^*(\bk,-\bp)
 +\beta_k^{0*}\alpha_p^0\cB(-\bk,\bp)
 +O(\varepsilon^2).
 \label{eq:deltaN}
\end{align}
These expressions make the cancellation of the first-order passive
contributions explicit.

The two covariances encode different momentum relations.
The anomalous covariance receives first-order contributions with
nonzero total momentum,
$\bk+\bp=\pm\bG_{\bn}$.
The normal covariance receives first-order contributions between
distinct momentum modes,
$\bk-\bp=\pm\bG_{\bn}$.
Thus, homogeneous evolution can distribute the correlations generated
by the medium between the normal and anomalous covariances.

The terms contributing to an individual covariance element may
interfere destructively. The invariant statement concerns the
complete covariance of the field and its canonical momentum,
as expressed by Eq.~\eqref{eq:fullcovtransfer}.
When each momentum block includes all components mixed by the
homogeneous evolution, the block is transformed by invertible
matrices on both sides. A nonzero block therefore remains nonzero.
This does not mean that its entries are unchanged or that every
element of $M^b$ and $N^b$ remains nonzero.
Nor does the persistence of such correlations establish
entanglement of a selected pair. That question must be answered
from the pair's reduced state.

For the ideal zero-average modulation, the
$\bG_{\bn}=0$ Fourier component is absent.
The first-order correction $\delta\beta_{\rm late}$ therefore
has no contribution supported at $\bp=-\bk$, where the
homogeneous pair-creation kernel is supported.
Consequently, the diagonal occupation of an exactly
momentum-resolved mode receives no first-order correction
from the modulation.
The same conclusion follows from Eq.~\eqref{eq:deltaN},
which has no contribution with $\bk-\bp=0$ at this order.

The complete second-order correction to the occupation requires
the second-order Bogoliubov transformation.
It contains both the contribution quadratic in
$\delta\beta_{\rm late}$ and the interference between the
zeroth-order kernel and the second-order correction.
Squaring Eq.~\eqref{eq:totalbeta} alone does not include the
latter contribution.

Finally, a pair-creation amplitude defined using the inflationary
mode basis and a particle occupation defined using the radiation-era
mode basis are different quantities.
They must be related through the full Bogoliubov transformation
before they can be compared.
For a specified momentum and helicity channel,
Eq.~\eqref{eq:totalbeta} can be approximated by
\[
 \delta\beta_{\rm late}(\bk,\bp)
 \simeq \alpha_k^0\cB(\bk,\bp)
\]
only when
\[
 \left|\beta_k^0\cA^*(-\bk,\bp)\right|
 \ll
 \left|\alpha_k^0\cB(\bk,\bp)\right|.
\]
Small $|\beta_k^0|$ alone is not sufficient, because the
pair-creation coefficient $\cB$ may itself be suppressed
or vanish in the channel under consideration.

Thus, the disappearance of the medium does not erase its reciprocal
correlation structure from the full covariance, but the survival
of entanglement in a selected pair remains a separate question.

\section{What survives as entanglement?}
\label{sec:survival}

Off-diagonal covariance is not itself an entanglement measure. A global unitary need not preserve entanglement across an arbitrary split of its oscillators. The relevant fact here is more specific. Homogeneous, helicity-preserving evolution mixes only $(\bk,\lambda)$ with $(-\bk,\lambda)$. For any partition that assigns both members of every such pair to the same side, the evolution factorizes as $U_A\otimes U_B$. Entanglement across that partition is unchanged, provided no modes are discarded and no environment is introduced.

For the equal-momentum channel, the helicity partition
\begin{equation}
 \{(\bk,R),(-\bk,R)\}\ \big|\ \{(\bk,L),(-\bk,L)\}
 \label{eq:protectedcut}
\end{equation}
is of this form in a closed four-mode sector. For the full field the corresponding split places all $R$ modes on one side and all $L$ modes on the other; Eq.~\eqref{eq:protectedcut} displays the four modes used below. The narrower split $(\bk,R)|(\bk,L)$, after tracing out both $-\bk$ modes, is not. This is the precise distinction between survival of the full helicity entanglement and survival of a co-propagating reduced pair. It does not require orientation averaging.

For the equal-momentum channel \(\mathbf G_{\mathbf n}=2\mathbf k\), the reality condition \(h_{-\mathbf n}=h_{\mathbf n}^{*}\) also supplies the channel \(-\mathbf G_{\mathbf n}=-2\mathbf k\). An isolated Gaussian completion that keeps these two channels is
\begin{equation}
 |\Psi_0\rangle=S_{+R,+L}(r,\theta_+)S_{-R,-L}(r,\theta_-)|0\rangle ,
 \label{eq:fourstate}
\end{equation}
where
\begin{equation}
S_{\pm R, \pm L}(r,\theta_\pm) 
= \exp \left[ r\left( e^{i\theta_\pm} a_{\pm R}^\dagger a_{\pm L}^\dagger
- e^{-i\theta_\pm} a_{\pm R} a_{\pm L} \right) \right] \ .
\end{equation}
Here \(+\) and \(-\) label four normalized discrete oscillators associated with momenta \(\mathbf k\) and \(-\mathbf k\), rather than delta-normalized continuum operators. We take \(r\ge0\), and expectation values are evaluated in \(|\Psi_0\rangle\).
At the exact co-propagating fixed point the odd center vertex vanishes, so this is the even sector of the combined model. This example omits other reciprocal partners and the higher-order effects of the pulses. It is an exactly solvable completion used to diagnose the later reduction, not the claimed exact output of the full EFT at finite $\varepsilon_E$.

For the equal-magnitude channels of the real pulse, the phases obey $\theta_++\theta_-=2\arg[i\cS_3(2k)]$ modulo $2\pi$. We allow general phases below so that the resulting bound also covers this relation. Put $n=\sinh^2r$, $m=\sinh r\cosh r$. Initially the only pair moments are
\begin{equation}
 M_{+R,+L}=m e^{i\theta_+},\qquad
 M_{-R,-L}=m e^{i\theta_-},\qquad
 N_{ii}=n.
 \label{eq:fourinitial}
\end{equation}
Let the same homogeneous map act on each helicity,
\begin{equation}
 b_{+\lambda}=\alpha_0a_{+\lambda}+\beta_0a_{-\lambda}^\dagger,
 \qquad |\alpha_0|^2-|\beta_0|^2=1.
 \label{eq:fourmap}
\end{equation}
Equations~\eqref{eq:Mtransform} and \eqref{eq:Ntransform}, now exact for this finite example, give the reduced $+$-direction moments
\(n_+=\langle b^\dagger_{+R}b_{+R}\rangle=\langle b^\dagger_{+L}b_{+L}\rangle\), \(m_+=\langle b_{+R}b_{+L}\rangle\) as
\begin{equation}
\begin{aligned}
 n_+=(|\alpha_0|^2+|\beta_0|^2)n+|\beta_0|^2\ ,\qquad
 m_+=m\bigl(\alpha_0^2e^{i\theta_+}+\beta_0^2e^{-i\theta_-}\bigr)\ .
\end{aligned}
 \label{eq:fourreduced}
\end{equation}
There is no cross-normal moment and no one-mode anomalous moment in this example. Define $q_\lambda=(b_{+\lambda}+b_{+\lambda}^{\dagger})/\sqrt2$ and $p_\lambda=(b_{+\lambda}-b_{+\lambda}^{\dagger})/(i\sqrt2)$, so that $[q_\lambda,p_\lambda]=i$. After local phase rotations, the symmetrized covariance of these quadratures is
\begin{equation}
 V_+=\begin{pmatrix}
 v&0&|m_+|&0\\0&v&0&-|m_+|\\|m_+|&0&v&0\\0&-|m_+|&0&v
 \end{pmatrix},\qquad v=n_++\frac12,
 \label{eq:fourcov}
\end{equation}
with quadratures ordered as $(q_R,p_R,q_L,p_L)$ and vacuum variance $1/2$.

Partial transposition changes $p_L$ to $-p_L$. The correlation block in $V_+$ then becomes $|m_+|I_2$. In the sum and difference coordinates $q_\pm=(q_R\pm q_L)/\sqrt2$ and $p_\pm=(p_R\pm p_L)/\sqrt2$, the transposed covariance has the two blocks $(v+|m_+|)I_2$ and $(v-|m_+|)I_2$. Its symplectic eigenvalues are $v\pm|m_+|$.
The partially transposed covariance is physical only if both symplectic eigenvalues are at least \(1/2\).  The smaller one is
\begin{equation}
 \widetilde\nu_-=n_++\frac12-|m_+|.
 \label{eq:PTeigen}
\end{equation}
For a two-mode Gaussian state the condition $\widetilde\nu_-<1/2$ is necessary and sufficient for entanglement \cite{Simon2000,Weedbrook2012}. Equivalently, the logarithmic negativity is
\begin{equation}
 E_{\mathcal N}^{(+)}=\max\{0,-\ln(2\widetilde\nu_-)\}.
 \label{eq:logneg}
\end{equation}
Thus a nonzero $m_+$ can coexist with a separable reduced pair if the added local noise is large enough.

Write $|\alpha_0|=\cosh s$ and $|\beta_0|=\sinh s$. The triangle inequality gives
\begin{equation}
 |m_+|\le\cosh(2s)m,\qquad
 \widetilde\nu_-\ge\frac12\cosh(2s)e^{-2r}.
 \label{eq:entbound}
\end{equation}
Equality is possible when the two contributions to $m_+$ have the same phase. Therefore
\begin{equation}
 \cosh(2s)e^{-2r}\ge1
 \quad\Longrightarrow\quad E_{\mathcal N}^{(+)}=0
 \label{eq:losscondition}
\end{equation}
for every phase choice in this example. 

At the same time, the homogeneous unitary factors between all $R$ and all $L$ modes. The total helicity entanglement in Eq.~\eqref{eq:protectedcut} is unchanged. Initially its entropy is $2[(n+1)\ln(n+1)-n\ln n]$ and its logarithmic negativity is $4r$; these values remain constant in the four-mode evolution. The example shows explicitly how entanglement can survive in an enlarged subsystem even when no entanglement remains between the two co-propagating output modes.

\section{Conclusions}\label{sec:conclusion}

We have formulated the parity-even kinetic and parity-odd spatial tensor couplings within a unified finite-time framework. The background determines the allowed total pair momenta, $\bk+\bp=\bG_{\bn}$, while the interaction operator determines the helicity weights. We have derived both the active and passive Bogoliubov kernels, including exact finite-angle contractions and closed-form time integrals for interaction pulses that end before radiation domination.

For co-propagating modes, the kinetic coupling creates opposite-helicity pairs with an amplitude that can remain nonzero when the two momenta are equal. When this equal-momentum pair is isolated from all other modes, the resulting state is a two-mode squeezed state of the $R$ and $L$ oscillators. For the chosen parity-odd operator, by contrast, the amplitude is proportional to $kp(p-k)$ and therefore vanishes at $k=p$.

After the interaction pulses end, the radiation-era evolution is spatially homogeneous, but the inherited correlations are not. The initially nonzero reciprocal blocks of the full covariance matrix of the field and canonical momentum remain nonzero under this invertible evolution. Nevertheless, the evolution can redistribute entanglement among the selected modes and their unobserved opposite-momentum partners. For a bipartition whose subsystems are each closed under this homogeneous evolution, including momentum reversal, the dynamics acts unitarily within each subsystem and preserves the entanglement between them. By contrast, entanglement in the reduced state of a co-propagating pair need not be preserved, as our four-mode example demonstrates.

Our main conclusions are threefold. Quasiperiodic order determines the allowed partner momenta. The interaction operators and their interference determine the helicity structure of the created pairs. Subsequent homogeneous evolution preserves the nonzero reciprocal covariance structure, but not necessarily the entanglement of a reduced pair.

Here, we have illustrated the entanglement dynamics using simple four-mode models. The full system contains a more complex network of correlations among momentum and helicity modes connected by the condition $\bk+\bp=\bG_{\bn}$. We leave a complete analysis of its entanglement structure for future work.
\vspace{0cm}
\section*{Acknowledgments}
 J.\ S. was in part supported by JSPS KAKENHI Grants No. JP23K22491, No. JP24K21548, and No. JP25H02186.

\appendix

\section{Polarization factors and the odd vertex}\label{app:polarization}
Choose $\hat\bk=(0,0,1)$ and $\hat\bp=(\sin\vartheta,0,\cos\vartheta)$. Transverse vectors are
\begin{equation}
 \bm t=(1,0,0),\quad\bm y=(0,1,0),\quad
 \bm t'=(\cos\vartheta,0,-\sin\vartheta).
 \label{eq:transverseframes}
\end{equation}
Use $\bm m_\sigma=(\bm t+i\sigma\bm y)/\sqrt2$ and $\bm m'_{\sigma'}=(\bm t'+i\sigma'\bm y)/\sqrt2$. Then
\begin{equation}
 \bm m_\sigma^*\cdot\bm m_{\sigma'}'{}^*
 =\frac{\cos\vartheta-\sigma\sigma'}2.
 \label{eq:vectoroverlap}
\end{equation}
Taking the square gives
\begin{equation}
 e^{\sigma*}:e^{\sigma'*}
 =\frac{(\cos\vartheta-\sigma\sigma')^2}{4}
 =\frac{(1-\sigma\sigma'\cos\vartheta)^2}{4}.
 \label{eq:tensoroverlap}
\end{equation}
This proves Eq.~\eqref{eq:angularP}. Expanding $\cos\vartheta=1-\vartheta^2/2+\vartheta^4/24+\cdots$ gives the near-collinear limits quoted in the text. Rotating either transverse frame only changes its helicity phase; probabilities and zeros do not depend on that choice.

The two-creation part of the odd spatial integrand contains
\begin{equation}
 (-k^2)(-ip_j)e_{i\ell}^{\lambda*}e_{k\ell}^{\lambda'*}
 =ik^2p_j e_{i\ell}^{\lambda*}e_{k\ell}^{\lambda'*}.
 \label{eq:creationderivatives}
\end{equation}
Symmetrizing it with the placement in which the two tensor legs are exchanged gives Eq.~\eqref{eq:Vplus}. Equation~\eqref{eq:helicityidentity} and its conjugate give
\begin{align}
 i\epsilon^{ijk}p_je_{k\ell}^{\lambda'}(\hat\bp)
 =\sigma_{\lambda'}p e_{i\ell}^{\lambda'}(\hat\bp)\ ,\qquad
 i\epsilon^{ijk}p_je_{k\ell}^{\lambda'*}(\hat\bp)
 =-\sigma_{\lambda'}p e_{i\ell}^{\lambda'*}(\hat\bp).\label{eq:curlcre}
\end{align}
The first symmetrized creation term is $-k^2p\sigma_{\lambda'}\cP^+/2$ and the second is $-p^2k\sigma_\lambda\cP^+/2$. The same calculation with one annihilation leg gives $\cP^-$ and the same scalar prefactor. This proves the two identities in Eq.~\eqref{eq:Vsimple}. It also verifies the symmetry and Hermiticity required by Eq.~\eqref{eq:firstcanonical}.

At a common direction, $\cP^+=\delta_{\lambda,-\lambda'}$. Thus the odd active vertex is
\begin{equation}
 \cV^+_{\lambda\lambda'}=-\frac{kp}{2}
 (k\sigma_{\lambda'}+p\sigma_\lambda)\delta_{\lambda,-\lambda'}.
 \label{eq:collappendix}
\end{equation}
For $RL$ this is $-kp(p-k)/2$; for $LR$ it is $+kp(p-k)/2$. Under exchange of both the momentum and helicity labels these two expressions agree, as required for bosons.

For a homogeneous coefficient, the annihilation-annihilation spatial vertex at opposite momenta is the complex conjugate of Eq.~\eqref{eq:Vplus} and equals $-\sigma k^3$ in a matched frame. The free spatial term in the action is $-a^2k^2|\gamma_\sigma|^2$. Adding $a c_O\chi(-\sigma k^3)|\gamma_\sigma|^2/\mpv$ gives Eq.~\eqref{eq:dispersion}. This check fixes its sign relative to the chosen action rather than leaving an independent sign choice in each equation.

\section{Radiation matching and covariance bookkeeping}\label{app:match}
At $\eta_1=-x/k$, multiply all canonical modes by $\sqrt{2k}$. The inflationary values in Eq.~\eqref{eq:vmode} are
\begin{align}
 \sqrt{2k}\,v_k^I(\eta_1)=(x^{-1}-i)e^{ix}\ , \qquad
 \frac{\sqrt{2k}}k v_k^{I\prime}(\eta_1)=(x^{-2}-1-ix^{-1})e^{ix}\ .
 \label{eq:vdstart}
\end{align}
For the radiation mode in Eq.~\eqref{eq:radmodes},
\begin{equation}
 \sqrt{2k}\,w_k(\eta_1)=-ie^{-ix},
 \qquad\frac{\sqrt{2k}}k w_k'(\eta_1)=-e^{-ix}.
 \label{eq:wstart}
\end{equation}
Put $A=\alpha_k^0e^{-ix}$ and $B=\beta_k^{0*}e^{ix}$. Continuity of the field and derivative gives
\begin{align}
 -iA+iB=(x^{-1}-i)e^{ix}\ ,\qquad
 -A-B=(x^{-2}-1-ix^{-1})e^{ix} \ .
 \label{eq:match2}
\end{align}
Adding and subtracting gives $B=-e^{ix}/(2x^2)$ and $A=(1+ix^{-1}-1/(2x^2))e^{ix}$. Equation~\eqref{eq:matchcoeff} follows. The normalization check is
\begin{equation}
 |\alpha_k^0|^2=\left(1-\frac1{2x^2}\right)^2+\frac1{x^2}
 =1+\frac1{4x^4}=1+|\beta_k^0|^2.
 \label{eq:matchnorm}
\end{equation}
If a kinetic coefficient instead jumped at the matching surface, the continuous canonical data would be the field and its projected momentum, not generally its unweighted derivative. Our interactions vanish before the background transition, so no such additional jump is present.

After switch-off the equations for field and momentum are
\begin{equation}
 \frac{\dd}{\dd\eta}\begin{pmatrix}\gamma_{\bk}\\\Pi_{\bk}\end{pmatrix}
 =\begin{pmatrix}0&4/(\mpl^2a^2)\\-\mpl^2a^2k^2/4&0\end{pmatrix}
 \begin{pmatrix}\gamma_{\bk}\\\Pi_{\bk}\end{pmatrix}.
 \label{eq:phaseflow}
\end{equation}
The generator has zero trace. If $T'=\mathsf L T$ and $T(\eta_s)=I$, then $\dd\ln\det T/\dd\eta=\tr\mathsf L=0$. Thus $\det T=1$, proving the invertibility used in Eq.~\eqref{eq:fullcovtransfer}. For any specified nonzero block, multiplication on the left and right by invertible matrices cannot turn it into the zero block.

The two operator maps also compose directly:
\begin{equation}
\begin{aligned}
 b_{\bk}={}&\alpha_k^0 a_{\bk}+\beta_k^0 a_{-\bk}^\dagger
+\alpha_k^0\intp[\cA(\bk,\bp)a_{\bp}+\cB(\bk,\bp)a_{\bp}^\dagger]\\
 &+\beta_k^0\intp[\cA^*(-\bk,\bp)a_{\bp}^\dagger+\cB^*(-\bk,\bp)a_{\bp}].
\end{aligned}
 \label{eq:mapexpanded}
\end{equation}
Reading off the coefficients proves Eqs.~\eqref{eq:totalalpha} and \eqref{eq:totalbeta}. Multiplying two such operators gives Eqs.~\eqref{eq:Mtransform} and \eqref{eq:Ntransform}, including the vacuum delta terms. In a purely passive initial transformation, $\cB=0$ and $\cA+\cA^\dagger=0$. All first-order changes of $M^b$ and $N^b$ cancel, because a passive transformation has left the initial vacuum unchanged. Keeping only $\delta\beta_{\rm late}$ would miss this cancellation.

\section{Partial transpose of the four-mode example}\label{app:PT}
For the selected forward pair, local phase rotations put its covariance in Eq.~\eqref{eq:fourcov}. Partial transpose in the $L$ subsystem reverses its momentum quadrature, so
\begin{equation}
 V_+^\Gamma=\begin{pmatrix}vI_2&|m_+|I_2\\|m_+|I_2&vI_2\end{pmatrix}.
 \label{eq:PTmatrix}
\end{equation}
The sum and difference coordinates separate it into $(v+|m_+|)I_2$ and $(v-|m_+|)I_2$. Its smallest symplectic eigenvalue is $\widetilde\nu_-=v-|m_+|$. The vacuum uncertainty bound is $1/2$, which gives Eq.~\eqref{eq:logneg} \cite{Simon2000,Weedbrook2012}.

Write $|\alpha_0|=\cosh s$, $|\beta_0|=\sinh s$, and $m=\sinh r\cosh r$. From Eq.~\eqref{eq:fourreduced},
\begin{align}
 n_++\tfrac12=\tfrac12\cosh(2s)\cosh(2r)\ ,\qquad
 |m_+|\le\tfrac12\cosh(2s)\sinh(2r)\ .\label{eq:mbound}
\end{align}
Subtracting proves
\begin{equation}
 \widetilde\nu_-\ge\tfrac12\cosh(2s)e^{-2r}.
 \label{eq:nuboundappendix}
\end{equation}
This bound applies to the reduced forward pair. It does not constrain the full $R|L$ entanglement in Eq.~\eqref{eq:protectedcut}, because homogeneous evolution is local with respect to that enlarged partition. The distinction is a choice of subsystem, not a failure of unitary evolution.

\bibliographystyle{unsrt}
\bibliography{pgw}

@article{Grishchuk1989,
  author = {Grishchuk, L. P. and Sidorov, Yu. V.},
  title = {On the Quantum State of Relic Gravitons},
  journal = {Class. Quantum Grav.},
  volume = {6},
  pages = {L161--L165},
  year = {1989},
  doi = {10.1088/0264-9381/6/9/002}
}

@article{Grishchuk1990,
  author = {Grishchuk, L. P. and Sidorov, Yu. V.},
  title = {Squeezed Quantum States of Relic Gravitons and Primordial Density Fluctuations},
  journal = {Phys. Rev. D},
  volume = {42},
  pages = {3413--3421},
  year = {1990},
  doi = {10.1103/PhysRevD.42.3413}
}

@article{Albrecht1994,
  author = {Albrecht, Andreas and Ferreira, Pedro and Joyce, Michael and Prokopec, Tomislav},
  title = {Inflation and Squeezed Quantum States},
  journal = {Phys. Rev. D},
  volume = {50},
  pages = {4807--4820},
  year = {1994},
  doi = {10.1103/PhysRevD.50.4807},
  eprint = {astro-ph/9303001},
  archivePrefix = {arXiv}
}

@article{AgulloBongaRibes2022,
  author = {Agullo, Ivan and Bonga, B{\'e}atrice and Ribes-Metidieri, Patricia},
  title = {Does Inflation Squeeze Cosmological Perturbations?},
  journal = {JCAP},
  volume = {2022},
  number = {09},
  pages = {032},
  year = {2022},
  doi = {10.1088/1475-7516/2022/09/032},
  eprint = {2203.07066},
  archivePrefix = {arXiv},
  primaryClass = {gr-qc}
}

@article{KannoSoda2022,
  author = {Kanno, Sugumi and Soda, Jiro},
  title = {Squeezed quantum states of graviton and axion in the universe},
  journal = {Int. J. Mod. Phys. D},
  volume = {31},
  number = {13},
  pages = {2250098},
  year = {2022},
  doi = {10.1142/S0218271822500985},
  eprint = {2112.14496},
  archivePrefix = {arXiv},
  primaryClass = {gr-qc}
}

@misc{SodaQC2026,
 author={Soda, Jiro},
 title={Quasicrystalline Inflation: From the Cosmological Dipole to Primordial Diffraction},
 year={2026}, eprint={2609.02142}, archivePrefix={arXiv}, primaryClass={hep-ph}}

@article{KangNicolis2016,
 author={Kang, J. and Nicolis, A.},
 title={Platonic Solids Back in the Sky: {Icosahedral} Inflation},
 journal={JCAP}, volume={2016}, number={03}, pages={050}, year={2016},
 doi={10.1088/1475-7516/2016/03/050},
 eprint={1509.02942}, archivePrefix={arXiv}, primaryClass={hep-th}}

@article{Lue1999,
 author={Lue, A. and Wang, L. and Kamionkowski, M.},
 title={Cosmological signature of new parity-violating interactions},
 journal={Phys. Rev. Lett.}, volume={83}, pages={1506}, year={1999},
 eprint={astro-ph/9812088}}

@article{AlexanderMartin2005,
 author={Alexander, S. and Martin, J.},
 title={Birefringent gravitational waves and the consistency check of inflation},
 journal={Phys. Rev. D}, volume={71}, pages={063526}, year={2005},
 eprint={hep-th/0410230}}

@article{SatohKannoSoda2008,
 author={Satoh, M. and Kanno, S. and Soda, J.},
 title={Circular polarization of primordial gravitational waves in string-inspired inflationary cosmology},
 journal={Phys. Rev. D}, volume={77}, pages={023526}, year={2008},
 eprint={0706.3585}, archivePrefix={arXiv}, primaryClass={astro-ph}}

@article{SatohSoda2008,
 author={Satoh, M. and Soda, J.},
 title={Higher curvature corrections to primordial fluctuations in slow-roll inflation},
 journal={JCAP}, volume={2008}, number={09}, pages={019}, year={2008},
 eprint={0806.4594}, archivePrefix={arXiv}, primaryClass={astro-ph}}

@article{TakahashiSoda2009,
 author={Takahashi, T. and Soda, J.},
 title={Chiral primordial gravitational waves from a {Lifshitz} point},
 journal={Phys. Rev. Lett.}, volume={102}, pages={231301}, year={2009},
 eprint={0904.0554}, archivePrefix={arXiv}, primaryClass={hep-th}}

@article{MaleknejadReview,
 author={Maleknejad, A. and Sheikh-Jabbari, M. M. and Soda, J.},
 title={Gauge fields and inflation},
 journal={Phys. Rept.}, volume={528}, pages={161}, year={2013},
 eprint={1212.2921}, archivePrefix={arXiv}, primaryClass={hep-th}}

@article{ObataSoda2016,
 author={Obata, I. and Soda, J.},
 title={Chiral primordial gravitational waves from dilaton induced delayed chromonatural inflation},
 journal={Phys. Rev. D}, volume={93}, pages={123502}, year={2016},
 eprint={1602.06024}, archivePrefix={arXiv}, primaryClass={hep-th}}

@article{AokiEFT2026,
 author={Aoki, K. and Fujita, T. and Kawaguchi, R. and Yanagihara, K.},
 title={Effective field theory of chiral gravitational waves},
 journal={JCAP}, volume={2026}, number={02}, pages={018}, year={2026},
 eprint={2504.19059}, archivePrefix={arXiv}, primaryClass={astro-ph.CO}}

@article{KannoDomain2023,
 author={Kanno, S. and Soda, J. and Taniguchi, A.},
 title={Circularly polarized gravitational waves in Chern-Simons gravity originating from an axion domain wall},
 journal={Phys. Rev. D}, volume={108}, pages={083525}, year={2023},
 eprint={2304.03944}, archivePrefix={arXiv}, primaryClass={hep-th}}

@article{SodaTakeuchi2026,
 author={Soda, Jiro and Takeuchi, Maki},
 title={Chiral gravitational waves from domain walls in {Nieh--Yan} gravity},
 journal={Phys. Rev. D}, volume={113}, pages={064044}, year={2026},
 doi={10.1103/mzd4-jhr4}, eprint={2512.14063}, archivePrefix={arXiv}, primaryClass={gr-qc}}

@article{Levine1984,
 author={Levine, D. and Steinhardt, P. J.},
 title={Quasicrystals: A New Class of Ordered Structures},
 journal={Phys. Rev. Lett.}, volume={53}, pages={2477--2480}, year={1984},
 doi={10.1103/PhysRevLett.53.2477}}

@article{Shechtman1984,
 author={Shechtman, D. and Blech, I. and Gratias, D. and Cahn, J. W.},
 title={Metallic phase with long-range orientational order and no translational symmetry},
 journal={Phys. Rev. Lett.}, volume={53}, pages={1951}, year={1984}}

@article{Baggioli2020,
 author={Baggioli, Matteo and Landry, Michael},
 title={Effective Field Theory for Quasicrystals and Phasons Dynamics},
 journal={SciPost Phys.}, volume={9}, pages={062}, year={2020},
 doi={10.21468/SciPostPhys.9.5.062},
 eprint={2008.05339}, archivePrefix={arXiv}, primaryClass={hep-th}}

@book{Humphreys1990,
  author = {Humphreys, James E.},
  title = {Reflection Groups and Coxeter Groups},
  series = {Cambridge Studies in Advanced Mathematics},
  volume = {29},
  publisher = {Cambridge University Press},
  address = {Cambridge},
  year = {1990},
  doi = {10.1017/CBO9780511623646}
}

@article{Cheung2008,
  author = {Cheung, Clifford and Creminelli, Paolo and Fitzpatrick, A. Liam and Kaplan, Jared and Senatore, Leonardo},
  title = {The Effective Field Theory of Inflation},
  journal = {JHEP},
  volume = {2008},
  number = {03},
  pages = {014},
  year = {2008},
  doi = {10.1088/1126-6708/2008/03/014},
  eprint = {0709.0293},
  archivePrefix = {arXiv},
  primaryClass = {hep-th}
}

@article{DeFeliceTsujikawa2015,
  author = {De Felice, Antonio and Tsujikawa, Shinji},
  title = {Inflationary Gravitational Waves in the Effective Field Theory of Modified Gravity},
  journal = {Phys. Rev. D},
  volume = {91},
  pages = {103506},
  year = {2015},
  doi = {10.1103/PhysRevD.91.103506},
  eprint = {1411.0736},
  archivePrefix = {arXiv},
  primaryClass = {hep-th}
}

@article{Nishizawa2018,
 author={Nishizawa, A. and Kobayashi, T.},
 title={Parity-violating gravity and GW170817},
 journal={Phys. Rev. D}, volume={98}, pages={124018}, year={2018},
 eprint={1809.00815}, archivePrefix={arXiv}}

@article{GaoHong2020,
 author={Gao, Xian and Hong, Xun-Yang},
 title={Propagation of the gravitational waves in a cosmological background},
 journal={Phys. Rev. D}, volume={101}, pages={064057}, year={2020},
 doi={10.1103/PhysRevD.101.064057},
 eprint={1906.07131}, archivePrefix={arXiv}, primaryClass={gr-qc}}

@misc{SongGao2026,
 author={Song, J.-J. and Gao, X.},
 title={Parity-violating spatially covariant gravity at total derivative order d=5},
 year={2026},
 eprint={2607.20317}, archivePrefix={arXiv}, primaryClass={gr-qc}}

@article{Weedbrook2012,
 author={Weedbrook, Christian and Pirandola, Stefano and Garc{\'i}a-Patr{\'o}n, Ra{\'u}l and Cerf, Nicolas J. and Ralph, Timothy C. and Shapiro, Jeffrey H. and Lloyd, Seth},
 title={Gaussian Quantum Information},
 journal={Rev. Mod. Phys.}, volume={84}, pages={621--669}, year={2012},
 doi={10.1103/RevModPhys.84.621},
 eprint={1110.3234}, archivePrefix={arXiv}, primaryClass={quant-ph}}

@article{Kothari2016,
 author={Kothari, Rahul and Ghosh, Shamik and Rath, Pranati K. and Kashyap, Gopal and Jain, Pankaj},
 title={Imprint of Inhomogeneous and Anisotropic Primordial Power Spectrum on {CMB} Polarization},
 journal={Mon. Not. R. Astron. Soc.}, volume={460}, number={2}, pages={1577--1587}, year={2016},
 doi={10.1093/mnras/stw1039},
 eprint={1503.08997}, archivePrefix={arXiv}, primaryClass={astro-ph.CO}}

@article{Simon2000,
 author={Simon, R.},
 title={{Peres--Horodecki} Separability Criterion for Continuous Variable Systems},
 journal={Phys. Rev. Lett.}, volume={84}, pages={2726--2729}, year={2000},
 doi={10.1103/PhysRevLett.84.2726},
 eprint={quant-ph/9909044}, archivePrefix={arXiv}}
\end{document}